\documentclass[reprint,aps,amsmath,amssymb,twoside,prd,showkeys,superscriptaddress]{revtex4-1}
\pdfoutput=1
\usepackage{verbatim}
\usepackage{comment}
\usepackage{graphicx}
\usepackage{graphicx}
\usepackage{epsfig}
\usepackage{bm}
\usepackage{tensor}
\usepackage{amsfonts}
\usepackage{amsmath}
\usepackage{graphicx}
\usepackage{hyperref}
\usepackage[english]{babel}
\usepackage[T1]{fontenc}
\usepackage[utf8]{inputenc}
\usepackage{booktabs}

\begin{document}
\title{Reconstructing the Type Ia Supernova Absolute Magnitude with Two-Probe Physics-Informed Neural Networks}

\author{Denitsa Staicova}
\email{dstaicova@inrne.bas.bg}
\affiliation{Institute for Nuclear Research and Nuclear Energy, Bulgarian Academy of Sciences}

\begin{abstract}
We apply two variants of Physics-Informed Neural Networks (PINNs) to
reconstruct the Type~Ia supernova absolute magnitude $M_B(z)$ from
joint BAO and supernova data under four cosmological models
($\Lambda$CDM, CPL, GEDE, $\Lambda_s$CDM) and two DESI~DR2 fiducial
sets. A heteroscedastic single-network method tested across four
constraint configurations establishes that the Etherington distance
duality relation is a more fundamental constraint than cosmological
model priors, reducing internal inconsistencies by up to an order of
magnitude. Under full constraints all models recover
$M_B \approx -19.3$~mag with biases below 0.05~mag. A Fisher
information-weighted two-network variant trains independent networks
on BAO and SN data, providing clean probe separation; it finds no
significant pointwise $M_B$ evolution in $z \in [0.3, 1.5]$, but
reveals a systematic separation of redshift-binned $M_B$ distributions.
The heteroscedastic method identifies a persistent $2$--$3\sigma$
residual at $z \sim 0.4$--$0.5$ that is consistent across all four
models and both fiducials, implying the same underlying tension.
While the origin of this feature remains ambiguous, its
model-independence and cross-method consistency warrant further
investigation with forthcoming data.
\end{abstract}

\maketitle
\section{Introduction}

The standardization of Type Ia supernovae (SNe~Ia) as cosmological distance indicators
rests on the assumption that their peak luminosities, after corrections for light curve
shape and color, yield a universal absolute magnitude $M_B \approx -19.3$~mag in the
$B$-band \cite{CosmoVerseNetwork:2025alb}. Accumulating evidence, however, suggests this assumption may be
oversimplified. Recent analyses report statistically significant evolution in Hubble
residuals that persist after accounting for observational systematics, attributed to
metallicity evolution in progenitor populations, changes in explosion physics, or
host-galaxy environmental dependencies correlated with redshift or as a sign of the Hubble tension \cite{Brout:2022vxf,Rose:2022zmu, Perivolaropoulos:2023iqj, Vagnozzi:2023nrq, Jiang:2024xnu, Barua:2024gei, DES:2024gpx, Yang:2024tkw, Thorp:2024rrs, Chung:2025gsy, vonMarttens:2025dvv,Rana:2025psr, DES:2022qsy}. While in some papers, there is a claim for more than $3\sigma$ deviation in $M_B(z)$, its position on the $z$-axis and physical origins differ. Rigorous, model-independent tests of $M_B(z)$ evolution are therefore needed for establishing the reliability of supernova-based cosmological constraints and for
identifying potential new physics.

Non-parametric reconstruction methods such as Gaussian Processes 
\cite{williams2006gaussian, Seikel:2012uu, Shafieloo:2012ht} and 
artificial neural networks \cite{Wang:2019vxv, Dialektopoulos:2021wde, 
Dialektopoulos:2023dhb} have been widely applied to cosmological 
distance reconstruction, offering model-independent alternatives to 
parametric approaches. In previous works we developed non-parametric reconstruction of
absolute magnitude $M_B(z)$ from BAO data using Gaussian Process (GP)
regression and artificial neural networks (ANN) \cite{Benisty:2022psx,
Staicova:2024dak}. The results showed that $M_B$ remains constant within
$1\sigma$ at current data precision, consistent with the analyses cited
above. In the current paper, we re-examine the reconstruction $M_B(z)$
through the PINN method and test whether it exhibits systematic evolution
beyond intrinsic supernova scatter.

The main tool we use is a Physics-Informed Neural Network (PINN) \cite{Raissi2019, raissi2024physicsinformedneuralnetworksextensions, Dai:2023kip, Bachar:2025eja, Verma:2025ujt} that
is capable of simultaneously fitting BAO and SNe~Ia data, while enforcing the Etherington distance duality relation $d_L = (1+z)^2 D_A$ and the model cosmology as hard loss term. We formulate training as heteroscedastic maximum likelihood --- where 
the network learns redshift-dependent uncertainties rather than assuming 
a fixed noise level --- learning both the distance-redshift relations 
and their uncertainties simultaneously. By enabling or disabling the distance duality
and cosmological model constraints independently, we obtain four configurations (TT,
TF, FT, FF) that isolate the contribution of each prior to the final $M_B(z)$
reconstruction. We apply this to four models spanning standard and dynamical dark
energy: $\Lambda$CDM, CPL \cite{Chevallier2001,Linder2003}, GEDE (Generalised Emergent Dark Energy)
\cite[][]{Li:2020ybr}, and LsCDM
\cite[Lambda-sign CDM,][]{Akarsu:2021fol, Akarsu:2024eoo}, using DESI~DR2 best-fit parameters
\cite{DESI:2025zgx} under two fiducial sets (DESI+CMB \cite{Planck2020} and DESI+PP \cite{Scolnic2022}). Additionally, we present results from an alternative version of the PINN code that instead uses the Fisher densities to sample the $z$ distribution and errors. The advantages of the Fisher information-weighted method are that  trains independent networks on BAO and SN data, providing clean probe separation.

\section{Theoretical Framework}

\subsection{Cosmological Distance Relations}

The angular diameter distance in a flat or curved FLRW cosmology is
\begin{equation}
D_A(z) = \frac{c}{H_0(1+z)} \int_0^z \frac{dz'}{E(z')} ,
\label{eq:DA}
\end{equation}
where $E(z) = H(z)/H_0$ and
$
E(z) = \sqrt{\Omega_m(1+z)^3 + \Omega_k(1+z)^2 + \Omega_\Lambda f_\text{DE}(z)} $. Since we use in flat cosmology, we use $\Omega_k=0$ and therefore $\Omega_{\Lambda} = 1 - \Omega_m$.

The dark energy evolution function $f_\text{DE}(z)$
equals unity for $\Lambda$CDM, $(1+z)^{3(1+w_0)}$ for a constant equation-of-state
model, and for the CPL parametrization \cite{Chevallier2001,Linder2003}
\begin{equation}
f_\text{DE}(z) = (1+z)^{3(1+w_0+w_a)} \exp\!\left(-\frac{3w_a z}{1+z}\right).
\end{equation}
The \textbf{(Generalised Emergent Dark Energy (GEDE)} model \cite{Li:2020ybr} introduces a
single parameter $\Delta$ that controls a smooth transition of the dark energy
density. The dimensionless Hubble parameter squared is same as above with $f_{DE}=f_\text{GEDE}$:

\begin{equation}
f_\text{GEDE}(z) = \frac{1 - \tanh\!\left[\Delta\log_{10}\!\left(\frac{1+z}{1+z_t}\right)\right]}
                        {1 + \tanh\!\left[\Delta\log_{10}(1+z_t)\right]}.
\end{equation}
The transition redshift $z_t$ is not a free parameter but is derived from the condition $\Omega_{DE}(z_t) = \Omega_m(1+z_t)^3$ (matter-dark energy equality).

For $\Delta = 0$, $f_\text{GEDE} = 1$ everywhere and
$\Lambda$CDM is recovered. For $\Delta < 0$ (the best-fit regime from DESI~DR2),
dark energy was suppressed at high redshift, easing the transition from matter
domination.

The \textbf{$\Lambda_s$CDM} (Lambda-sign CDM) model \cite{Akarsu:2021fol,Akarsu:2024eoo}
extends $\Lambda$CDM by allowing the cosmological constant to switch sign at a
transition redshift $z_\dagger$, motivated by a conjectured transition from
anti-de~Sitter to de~Sitter vacua in the late universe. The original model defines
an abrupt sign switch $\Lambda_s \propto \mathrm{sgn}(z_\dagger - z)$, giving
\begin{equation}
E^2(z) = \Omega_m(1+z)^3 + \Omega_\Lambda\,\mathrm{sgn}(z_\dagger - z),
\end{equation}
where $\Omega_\Lambda = 1 - \Omega_m$. For $z < z_\dagger$ the cosmological term is
positive (de~Sitter phase, consistent with the observed accelerated expansion); for
$z > z_\dagger$ it is negative (anti-de~Sitter phase). Since a true step function is
not differentiable and therefore incompatible with gradient-based PINN training, we
replace the sign function with a smooth approximation:
\begin{equation}
E^2(z) = \Omega_m(1+z)^3 + \Omega_\Lambda\,\tanh\!\left[\eta\,(z_\dagger - z)\right],
\label{eq:lscdm}
\end{equation}
with $\eta = 10^{1.5} \approx 31.6$, which reproduces the step function to within
$10^{-14}$ everywhere except within $\Delta z \sim 0.1$ of $z_\dagger$. The model
predicts a local minimum in $E(z)$ near $z_\dagger$ and a modified expansion history
relative to $\Lambda$CDM, with $\Lambda$CDM recovered only in the limit
$z_\dagger \to \infty$.

The Etherington distance duality relation (DDR), connecting the angular and luminosity distance:
\begin{equation}
d_L(z) = (1+z)^2 D_A(z),
\label{eq:duality}
\end{equation}
holds in any metric theory of gravity with photon number conservation. The distance
modulus is
\begin{equation}
\mu(z) = 5\log_{10}(d_L(z)/\text{Mpc}) + 25 ,
\label{eq:mu}
\end{equation}
and the absolute magnitude enters as $\mu_\text{obs} = \mu(z) + M_B$ for the
standardized SN~Ia apparent magnitude.

\subsection{Observational Data}
\label{subsec:Data}
BAO surveys measure $D_M(z)/r_d = (1+z)D_A(z)/r_d$, where $r_d$ is the sound horizon
at the baryon drag epoch. We fix $r_d = 147.0$~Mpc consistent with Planck~2018
\cite{Planck2020}; as shown in \cite{Benisty:2022psx}, alternative values shift $M_B$ by a constant offset without affecting the redshift-dependent reconstruction of $M_B(z)$. We use measurements from DESI~DR2 \cite{DESI:2025zgx} spanning
$0.3 \lesssim z \lesssim 2.33$.

For supernovae we use the Pantheon+ compilation \cite{Brout:2022vxf, Scolnic2022}
($N \approx 1700$ SNe~Ia, $0.01 < z < 2.3$) with its full covariance matrix; We adopt the
fiducial $M_B = -19.3$~mag noting that the Planck-preferred
value ($-19.44$~mag) would shift absolute biases by $\sim 0.14$~mag but not affect
relative comparisons. We use two fiducial models: under two fiducial sets (DESI+CMB \cite{Planck2020} and DESI+PP \cite{Scolnic2022}) to check how much the results depend on the fiducial cosmology.

Table~\ref{tab:models} lists the four fiducial models and their DESI~DR2 best-fit parameters
under the two fiducial sets used in this work with values from \cite{DESI:2025zgx} .

\section{PINN Architecture and Training}
\label{sec:pinn}

We apply two variants of Physics-Informed Neural Networks to reconstruct
$M_B(z)$: a single shared-backbone heteroscedastic network that jointly
fits BAO and supernova data with physical constraints, and a Fisher
information-weighted variant using two fully independent networks --- one
trained on BAO data, one on supernovae --- that provides clean separation
of the two probes.

\subsection{Heteroscedastic Method}
\label{sec:heteroscedastic}

\subsubsection{Network Design}
\label{sec:network}

We use a feed-forward network mapping $z \to \{D_A/r_d,\, \mu,\,
\log\sigma^2_{D_A},\, \log\sigma^2_\mu\}$, with four hidden layers
(128--64--64--32 neurons) and Swish activations. The input is normalized as
$\tilde{z} = z/3$. The two distance outputs use softplus activations with
physically motivated bias initialization ($D_A/r_d$: bias 10.0; $\mu$: linear
with bias 40.0). The log-variance outputs convert to standard deviations as
$\sigma = \exp(s/2)$ at inference time, where $s$ is the raw network output.
This log-variance parametrization is numerically stable and naturally aligned
with the negative log-likelihood loss described below.

\subsubsection{Loss Function}
\label{sec:hetero_loss}

The total loss is
\begin{equation}
\mathcal{L} = \mathcal{L}_\text{NLL-BAO} + \mathcal{L}_\text{NLL-SN}
            + \lambda_\text{phys}\,\mathcal{L}_\text{phys}
            + \lambda_\text{cosmo}\,\mathcal{L}_\text{cosmo} ,
\label{eq:loss_total}
\end{equation}
where each term is described below.

\paragraph{Data terms.}
The heteroscedastic NLL for observable $y$ with known measurement uncertainty
$\sigma_\text{obs}$ and learned model uncertainty $\sigma_\text{model}(z)$ is
\begin{align}
&\mathcal{L}_\text{NLL} =\nonumber \\ &\frac{1}{2N}\sum_{i=1}^N
\left[\frac{(y_i^\text{obs} - y_i^\text{pred})^2}{\sigma^2_{\text{obs},i} + e^{s_i}}
+ \log\!\left(\sigma^2_{\text{obs},i} + e^{s_i}\right)
- \log\sigma^2_{\text{obs},i}\right],
\label{eq:loss_nll}
\end{align}
where $s_i = \log\sigma^2_{\text{model},i}$ is the network output. This form
is applied to both BAO and SN data. For the supernova term the diagonal
$\sigma^2_\text{obs}$ in the fit residual is replaced by the full
Pantheon+ covariance matrix $\mathbf{C}$ (or the PCA equivalent), and the NLL uncertainty term is
down-weighted by $0.1$ to prevent variance learning from dominating fit quality.
A known limitation of this weighting is that the network tends to learn
$\sigma_\text{model} \approx 0$; we address this with an excess-variance
parametrisation in Appendix~\ref{app:conservative_errors} and show the
main conclusions are robust.

\paragraph{Physics constraint (DDR).}
At $N_\text{coll} = 500$ collocation points $z_i \sim \mathcal{U}(0,\,2.5)$
resampled each iteration:
\begin{equation}
\mathcal{L}_\text{phys} = \frac{1}{N_\text{coll}}\sum_{i=1}^{N_\text{coll}}
\left[\mu_i^\text{pred} - \mu_i^\text{DDR}\right]^2,
\label{eq:loss_phys}
\end{equation}
where $\mu^\text{DDR} = 5\log_{10}[(1+z)^2 D_A^\text{pred} r_d] + 25 + M_B$.
This enforces internal consistency between the two network output heads
across the full redshift range.

\paragraph{Cosmological model constraint.}
\begin{align}
&\mathcal{L}_\text{cosmo} = \frac{1}{N_\text{coll}} \times
\nonumber \\
&\sum_{i=1}^{N_\text{coll}}
\left[(D_A/r_d)_i^\text{pred} - (D_A/r_d)_i^\text{theory}\right]^2
+ \left[\mu_i^\text{pred} - \mu_i^\text{theory}\right]^2,
\label{eq:loss_cosmo}
\end{align}
where both $(D_A/r_d)^\text{theory}$ and $\mu^\text{theory}$ are evaluated from
the fiducial cosmological model Eq.~(\ref{eq:DA}).

\paragraph{Constraint configurations.}
Setting $\lambda_\text{phys} \in \{0, 10\}$ and $\lambda_\text{cosmo} \in \{0, 30\}$
independently yields four configurations: \textbf{TT} (both enabled), \textbf{TF}
(DDR only), \textbf{FT} (cosmology only), \textbf{FF} (purely data-driven). These
allow systematic isolation of each prior's contribution to the reconstruction.

\subsection{Fisher Information-Weighted Extension}
\label{sec:fisher}

Standard heteroscedastic training weights data by $\sigma_i^{-2}$ but does not account
for non-uniform redshift sampling. Dense low-$z$ coverage can cause the network to
overfit the crowded region while underfitting sparse high-$z$ data. To address this, we
developed a Fisher information-weighted variant that incorporates data quality and
spatial density jointly, and separates the supernova and BAO inference channels into
independent networks to obtain a clean $M_B(z)$ reconstruction. This separation enforces statistical independence between the BAO and SN inference channels, preventing implicit coupling through a shared representation and ensuring that any reconstructed signal in $M_B(z)$ reflects a genuine tension between the two probes rather than internal consistency of a joint fit.

\subsubsection{Two-Network Architecture}
\label{sec:twonet}

The heteroscedastic method of Section~\ref{sec:heteroscedastic} uses a single
shared-backbone network that outputs both $D_A/r_d(z)$ and $\mu(z)$, which entangles
the two probes and obscures the errors of the residual $M_B(z)$. The Fisher variant replaces the single network with two fully independent
heteroscedastic networks trained with separate optimisers and gradient tapes:
\begin{itemize}
\item \textbf{DA-net}: $z \mapsto \bigl(D_A/r_d,\;\log\sigma^2_{D_A}\bigr)$, trained
  only on BAO data plus physics regularisation (DDR consistency, derivative matching,
  boundary conditions).
\item \textbf{$\mu$-net}: $z \mapsto \bigl(\mu,\;\log\sigma^2_\mu\bigr)$, trained
  on supernova data only (with the full PCA-compressed covariance).
\end{itemize}
Both networks share the same architecture ($[32, 128, 64, 64, 32]$ with tanh
activations) but have entirely independent weights. The absolute magnitude is then
reconstructed as
\begin{equation}
M_B(z) = \mu_\mathrm{net}(z)
        - 5\log_{10}\!\bigl[(1{+}z)^2\,D_{A,\mathrm{net}}(z)\,r_d\bigr] - 25 ,
\label{eq:MB_twonet}
\end{equation}
and because the two networks are statistically independent, the uncertainty adds in
quadrature:
\begin{equation}
\sigma^2_{M_B}(z) = \sigma^2_{\mu,\mathrm{net}}(z)
    + \left(\frac{\partial\mu}{\partial D_A}\right)^{\!2}\sigma^2_{D_A,\mathrm{net}}(z) ,\label{eq:sigma_MB}
\end{equation}

where $
\frac{\partial\mu}{\partial D_A} = \frac{5}{\ln 10}\,\frac{1}{D_A/r_d} $.

This construction ensures that any signal in $M_B(z)$ reflects a genuine tension
between the two datasets rather than internal self-consistency of a single fit.
\subsubsection{Loss Function Structure}
\label{sec:fisher_loss}

Both networks use the excess-variance parametrisation
$\sigma^2_\mathrm{total} = \sigma^2_\mathrm{obs}(1 + e^s)$,
ensuring $\sigma_\mathrm{total} \geq \sigma_\mathrm{obs}$.
The variance NLL per point is
\begin{equation}
\ell_\mathrm{var}(r_i, s_i) = \frac{1}{2}\left[
  \frac{r_i^2}{\sigma^2_{\mathrm{obs},i}(1+e^{s_i})}
  + \log(1 + e^{s_i})\right],
\label{eq:lvar}
\end{equation}
where $r_i$ is the residual. The DA-net loss combines data, physics, and
regularisation terms:

\begin{align}
&\mathcal{L}_\mathrm{DA} = \nonumber\\
 & \underbrace{\sum_i w_i \mathcal{L}_{\chi^2}^\mathrm{BAO}
    + \sum_i w_{\mathrm{nll},i}\,\ell_\mathrm{var}}_\text{BAO data}
  + \nonumber \\
&  \underbrace{\lambda_\mathrm{phys}\,\mathcal{L}_\mathrm{DDR}
    + \lambda_\mathrm{deriv}\,\mathcal{L}_{dD_A/dz}}_\text{physics}
  + \mathcal{L}_\mathrm{reg}^{D_A},
  \label{eq:loss_DA}
\end{align}

Where $ \mathcal{L}_{\chi^2}^\mathrm{BAO} = {r_i^2}/{\sigma^2_{\mathrm{obs},i}}$ is the standard $\chi^2$ terms comparing measurements with their observational errors, with hybrid
Fisher density weights $w_i$ concentrating the gradient on
the most informative data points. The second term is the variance
NLL weighted by inverse-Fisher weights $w_{\mathrm{nll},i}$,
learning the redshift-dependent uncertainty. Both weight sets are
defined in Section~\ref{sec:fisher_weights}. The derivative loss $\mathcal{L}_{dD_A/dz}$ penalises deviations of the
network slope $\mathrm{d}D_A/\mathrm{d}z$ from the theoretical prediction via
finite differences, encoding the cosmological model's shape constraint. The regularization terms in both cases are described in Appendix \ref{app:implementation}.

The $\mu$-net is purely data-driven:
\begin{equation}
\mathcal{L}_\mu =
  \underbrace{\mathcal{L}_{\chi^2}^\mathrm{SN}
    + 5\sum_i w_{\mathrm{nll},i}\,\ell_\mathrm{var}}_\text{SN data}
  + \mathcal{L}_\mathrm{reg}^{\mu},
  \label{eq:loss_mu}
\end{equation}
with $\mathcal{L}_{\chi^2}^\mathrm{SN}$ computed using the full PCA-compressed
Pantheon+ covariance. The factor of 5 on the variance NLL balances the large
number of supernovae against the BAO chi2 scale. No physics prior enters
$\mathcal{L}_\mu$, ensuring the supernova distance modulus is a
model-independent reconstruction.

\subsubsection{Fisher Information Weighting}
\label{sec:fisher_weights}

Standard heteroscedastic NLL training weights data implicitly by
$\sigma_i^{-2}$ but ignores non-uniform redshift sampling. To address both
simultaneously we construct two sets of weights from the per-point Fisher
information
\begin{equation}
\mathcal{I}_i = \frac{1}{\sigma_i^2}\sum_\theta
  \left(\frac{\partial f}{\partial\theta}\right)^{\!2},
\label{eq:fisher_info}
\end{equation}
where the sum runs over all active cosmological parameters $\theta$ and
derivatives are evaluated by finite differences at the fiducial values.

\textbf{Hybrid $\chi^2$ weights.} The mean-network gradient is concentrated on
the most informative data points via
\begin{equation}
w_i = \mathcal{I}_i^{\,\alpha_F} \cdot
      \frac{1}{\sqrt{N_\text{neighbors}(z_i)}},
\label{eq:hybrid_weights}
\end{equation}
where $N_\text{neighbors}(z_i)$ counts points within a fixed redshift
bandwidth (inverse square-root density weighting) and we fix $\alpha_F =
0.5$.These weights $w_i$ enter the $\chi^2$ term of the DA-net loss
(Eq.~\ref{eq:loss_DA}).

\textbf{Inverse-Fisher NLL weights.} The variance head of each network receives
gradient signal only at observed data points, giving it no direct information
about uncertainty in sparse redshift regions. To address this, the variance NLL
term is weighted by the \emph{inverse} Fisher information, interpolated from a
combined BAO$+$SN Fisher curve and floored at its 5th percentile:
\begin{equation}
w_{\mathrm{nll},i} = \frac{1}{\mathcal{I}(z_i)},
\quad \text{normalised to } \max_i(w_{\mathrm{nll},i}) = 1,
\quad \text{floor } 0.1 .
\label{eq:inv_fisher_nll}
\end{equation}
Where Fisher information is low (sparse data, large errors), the NLL weight is
high, penalising underestimated uncertainty more severely. Separate weight
arrays $w_{\mathrm{nll}}^\mathrm{BAO}$ and $w_{\mathrm{nll}}^\mathrm{SN}$
are evaluated at the BAO and SN redshifts respectively and passed to the
DA-net and $\mu$-net variance losses.

\subsection{Collocation Sampling and Training}
\label{sec:collocation}

Physics collocation points for the heteroscedastic method are sampled
uniformly from $z \in [0.001, 2.5]$ and resampled at each training step.
For the Fisher variant, points are drawn from a mixture of 50\% uniform
and 50\% adaptive samples from $p(z) \propto \mathcal{I}_\mathrm{total}(z)^\beta$
with $\beta = +1$, supplemented by the BAO redshifts themselves, so that
physics enforcement is concentrated in information-rich regions. The physics
loss on collocation points is additionally weighted by inverse Fisher
information. Both methods use a four-phase training schedule with progressive
constraint ramping; full details are given in Appendix~\ref{app:implementation}.

\subsection{Uncertainty Quantification}
\label{sec:uncertainty}

The PINN framework produces two distinct uncertainty measures on $M_B(z)$
that we use throughout the paper. The \emph{per-point predictive uncertainty}
\begin{equation}
\sigma^2_{M_B}(z) = \sigma^2_{\mu,\mathrm{model}}
  + \left(\frac{5}{\ln 10}\right)^2 \frac{\sigma^2_{D_A,\mathrm{model}}}{D_A^2(z)} ,
\label{eq:sigmaMB}
\end{equation}
propagates the two learned variances through the distance duality relation and
represents the \emph{aleatoric} uncertainty of the fit at redshift $z$, as opposed to the \emph{epistemic} uncertainty of a GP reconstruction  which quantifies how well the function is constrained given finite, noisy data. A theoretical
lower bound follows from the observational inputs: the BAO term contributes
$\sim 0.02$--$0.05$~mag and adding the SN measurement floor $\sigma_\mu \sim
0.1$~mag in quadrature gives a physical floor of $\sim 0.1$~mag.

The \emph{distribution scatter}
\begin{equation}
\sigma^\mathrm{dist}_{M_B} = \mathrm{std}_z\bigl[M_B(z)\bigr]
\end{equation}
measures how much the inferred absolute magnitude varies across
cosmic time for a given cosmological model --- essentially the
amplitude of any redshift trend in $M_B(z)$, independently of the
per-point uncertainty. A statistically significant non-zero value
would indicate either a genuine physical evolution of the SN~Ia
population or a tension between the assumed cosmology and the data.
A value $\sigma^\mathrm{dist}_{M_B} \ll 0.1$~mag is consistent with
a redshift-independent $M_B$.

A known limitation of both PINN variants is that the variance head receives
gradient signal only at observed data points, so it has no direct information
about uncertainty in unsampled redshift regions. In the standard heteroscedastic
method this manifests as uncertainties that remain roughly flat or even decrease
toward high $z$, contrary to the expectation that sparse coverage should imply
larger uncertainty. The Fisher inverse-weighting partially addresses this by
penalising the variance head more severely where Fisher information is low,
and the effect is visible in Fig.~\ref{fig:fisher_delta_MB}: Fisher
uncertainties grow appropriately at $z < 0.2$ and $z > 1.5$, while the
heteroscedastic uncertainties remain underestimated in those regions.

By contrast, Gaussian process reconstructions automatically inflate uncertainty
away from data points through the kernel length-scale, which superficially
looks more conservative. However, this inflation is itself a prior assumption
— a GP with a long kernel will report small uncertainty even in gaps if the
surrounding data is precise, while a short kernel inflates uncertainty
aggressively regardless of data quality. Neither approach is strictly superior;
the PINN uncertainty is more honest about what the data directly constrain,
while the GP uncertainty is more honest about what the model cannot exclude. We compare our results quantitatively to the GP reconstruction of \cite{Benisty:2022psx} in Section~\ref{sec:discussion_gp}.

\label{sec:results}
\begin{figure*}[ht]
\centering
\includegraphics[width=\textwidth]{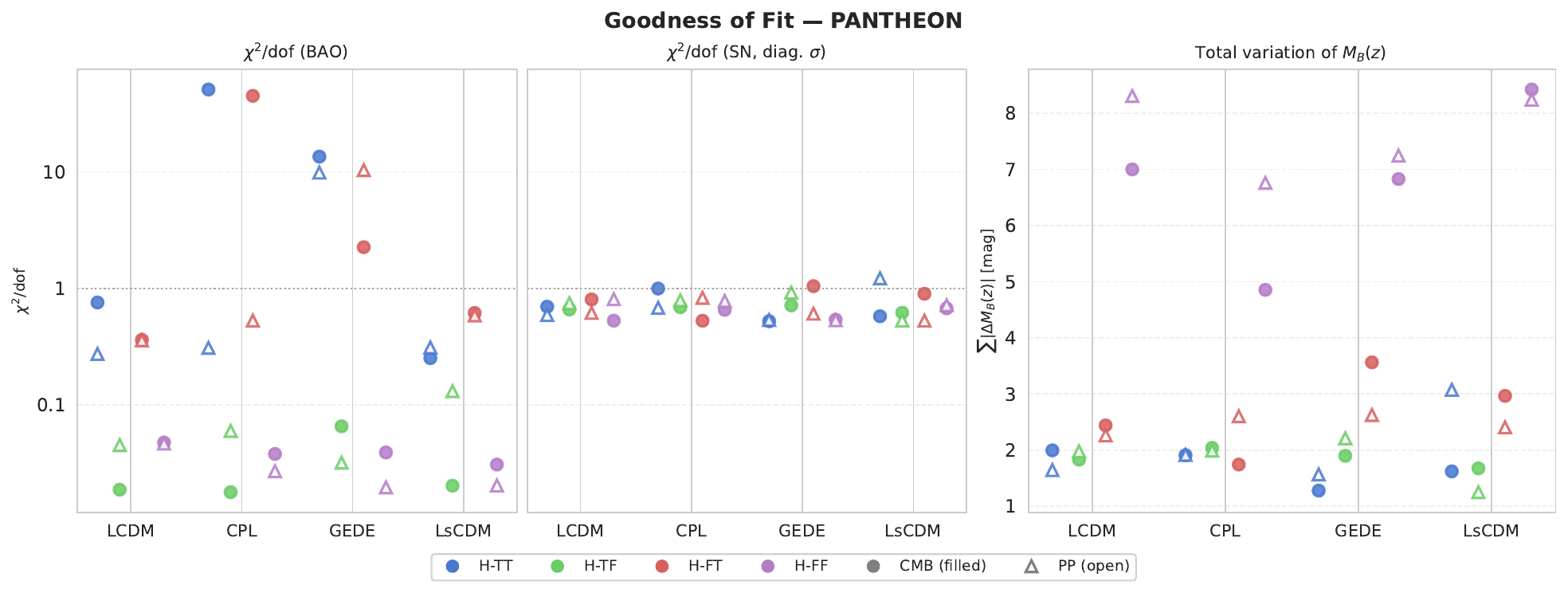}
\caption{Goodness-of-fit summary for all models, configurations, and fiducials on
the Pantheon+ dataset. \emph{Left and centre}: $\chi^2/\mathrm{dof}$ for BAO and
SN~Ia (diagonal errors), respectively, on a shared logarithmic scale. The dotted
line marks $\chi^2/\mathrm{dof} = 1$. \emph{Right}: total variation
$\sum|\Delta M_B(z)|$ of the reconstructed $M_B(z)$ curve, a measure of
high-frequency oscillation. Filled circles indicate DESI+CMB fiducial; open triangles
indicate DESI+PP. Colours denote the constraint configuration: H-TT (blue), H-TF
(green), H-FT (red), H-FF (purple). See text for a discussion on the outliers }
\label{fig:gof_summary}
\end{figure*}

\subsection{$M_B$ Reconstruction}
\label{sec:mb_reconstruction}

At each test redshift the absolute magnitude is extracted as
\begin{equation}
M_B(z) = \mu^\text{pred}(z)
        - 5\log_{10}\!\left[(1+z)^2\,D_A^\text{pred}(z)\,r_d\right] - 25 ,
\label{eq:MB_extraction}
\end{equation}
with uncertainty propagated as in Eq.~(\ref{eq:sigmaMB}). For the Fisher
variant, $\mu^\text{pred}$ and $D_A^\text{pred}$ come from the independent
$\mu$-net and DA-net respectively, so any signal in $M_B(z)$ reflects a
genuine tension between the two datasets. We evaluate
Eq.~(\ref{eq:MB_extraction}) on a uniform grid of 200 redshifts spanning
$z \in [0.001,\, 2.5]$.

\section{Results}

We trained PINNs for all combinations of four models
($\Lambda$CDM, CPL, GEDE, LsCDM), two fiducial sets (DESI+CMB, DESI+PP), and for the heteroscedastic case on four
constraint configurations (TT, TF, FT, FF). Primary analysis uses Pantheon+ and DESI DR2 datasets.

\subsection{Heteroscedastic Method}
\label{sec:hetero_results}

\textbf{Goodness of Fit} Figure~\ref{fig:gof_summary} summarises the goodness of fit across all
models, constraint configurations, and fiducials on the Pantheon+ dataset.
BAO and SN $\chi^2/\mathrm{dof}$ values cluster near unity for the
constrained configurations (H-TT, H-TF), confirming adequate fits to both
probes simultaneously. The notable exception is CPL under the DESI+CMB
fiducial, which shows catastrophically poor BAO fits
($\chi^2/\mathrm{dof} \sim 50$), reflecting the known tension between
CMB-preferred CPL parameters and BAO data. The DESI+PP fiducial resolves this tension for CPL, consistent
with the better mutual consistency of the PP-preferred parameters.

The total variation $\sum|\Delta M_B(z)|$ in the right panel of
Fig.~\ref{fig:gof_summary} provides a diagnostic for unphysical
oscillations: unconstrained configurations (H-FF) reach $\sim 7$--8~mag,
while constrained methods cluster at $\sim 1.5$--2~mag, consistent with
genuine data features rather than network freedom.

\textbf{DDR violation:} We quantify internal consistency between the two network output heads via
the DDR violation $\Delta_\mathrm{DDR}(z) = |\mu^\mathrm{pred}(z) -
\mu^\mathrm{DDR}(z)|$, where $\mu^\mathrm{DDR} = 5\log_{10}[(1+z)^2
D_A^\mathrm{pred} r_d] + 25 + M_B$.
A well-trained and DDR-consistent model should have
$|\Delta_\mathrm{DDR}| < 50$~mmag across $0 < z < 2.5$.

Table~\ref{tab:ddr_summary} quantifies DDR compliance across all runs.
TT and TF configurations achieve mean violations of 30--52~mmag,
while FT and FF show violations 2--50$\times$ larger. This establishes
a clear hierarchy: the Etherington distance duality relation provides
more fundamental regularisation than the specific cosmological model
prior.

\begin{figure*}[ht]
\centering
\includegraphics[width=\textwidth]{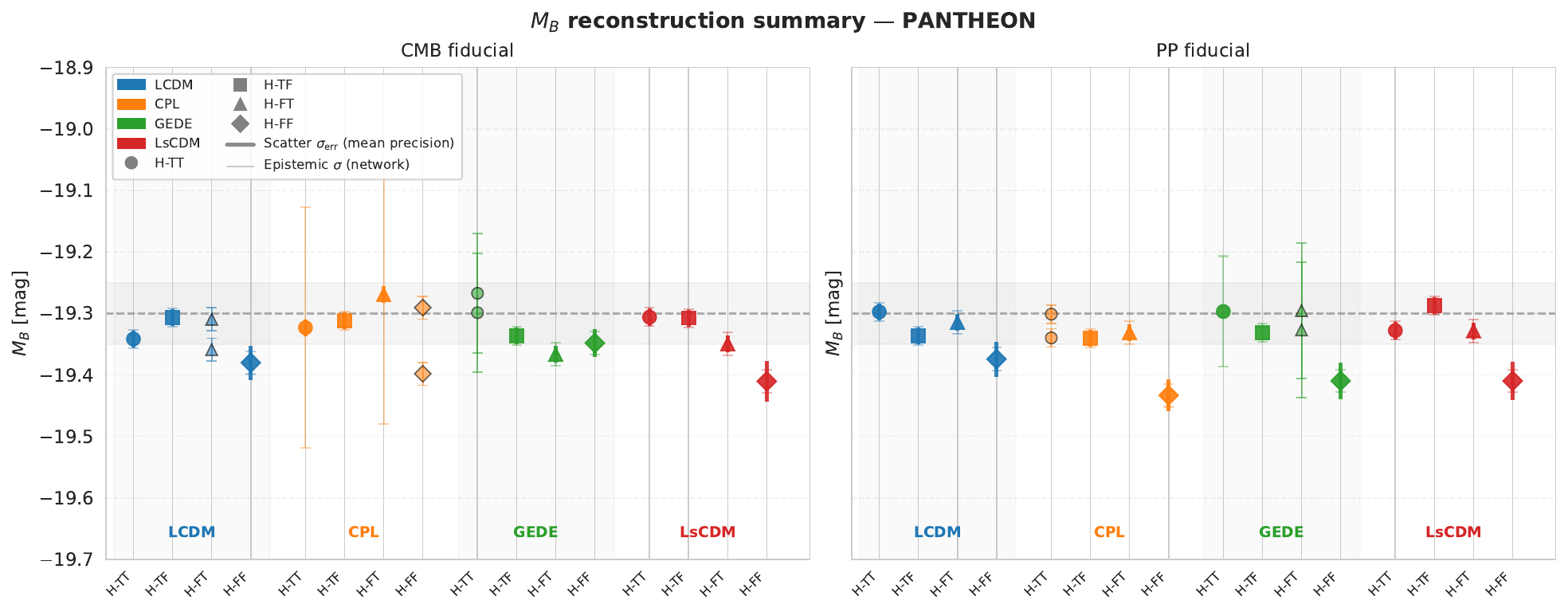}
\caption{Summary of reconstructed mean $M_B$ for all models, configurations, and
fiducials on the Pantheon+ dataset. Left panel shows DESI+CMB, right: DESI+PP. Colours denote the cosmological model; marker shapes
denote the constraint configuration. Thick error bars show the scatter-based standard error on the mean
$\sigma_\mathrm{err}$; thin error bars show the epistemic uncertainty from the network
$\sigma_\mathrm{ep}$. The dashed line and the shaded band marks $M_{B,\mathrm{fid}} = -19.3 \pm 0.5$ mag.
}
\label{fig:mb_summary}
\end{figure*}

\begin{table}[h]
\centering

\begin{tabular}{lcccc}
\toprule
Config & \multicolumn{2}{c}{DESI+CMB} & \multicolumn{2}{c}{DESI+PP} \\
       & Mean & Max & Mean & Max \\
\hline
TT & 31--51 & 120--193 & 32--46 & 109--182 \\
TF & 36--52 & 134--193 & 32--51 & 115--179 \\
FT & 85--181 & 276--449 & 85--195 & 302--857 \\
FF & 120--579 & 250--5115 & 127--178 & 256--664 \\
\hline
\end{tabular}
\caption{DDR violations [mmag], Pantheon+ dataset. Each cell shows the range of mean and max values across four cosmological models.}
\label{tab:ddr_summary}
\end{table}

\label{sec:mb_results}

\textbf{$M_B$ Reconstruction:}
Table~\ref{tab:mb_hetero} and Fig.~\ref{fig:mb_summary} present the
reconstructed $M_B$ for all models, fiducials, and constraint configurations
on the Pantheon+ dataset. Under full physical constraints (H-TT) all four
models recover $M_B$ consistent with constancy at $M_B \approx -19.3$~mag,
with values spanning $-19.27$ to $-19.34$~mag across all models and both
fiducials. The DESI+PP fiducial generally yields values closer to $-19.3$
than DESI+CMB, consistent with the better BAO fit quality seen in
Fig.~\ref{fig:gof_summary}.

\textbf{Bimodality} found with kernel density estimation with peak-finding, appears in several configurations and is not confined to
unconstrained runs as might be naively expected. GEDE shows bimodality
already under H-TT (CMB fiducial), CPL under H-TT (PP fiducial) and H-FF
(CMB fiducial), and $\Lambda_s$CDM is unimodal in all configurations.
$\Lambda$CDM shows bimodality only under H-FT. We interpret this pattern
as reflecting genuine degeneracy in the $D_A$--$\mu$ loss landscape for
dynamical dark energy models: without both the DDR and cosmological model
constraints simultaneously, the network can trade off $D_A$ and $\mu$ in
two distinct ways that fit the data equally well. The fact that this
degeneracy persists even under H-TT for GEDE and CPL suggests it is
driven by the additional freedom in $w(z)$ rather than by insufficient
regularisation. $\Lambda$CDM and $\Lambda_s$CDM, whose expansion histories
are more tightly constrained, remain unimodal under full constraints. All reconstructed values are within $\sim 0.15$~mag of the fiducial
$M_{B,\mathrm{fid}} = -19.3$~mag regardless of configuration, confirming
that the PINN does not produce pathological solutions.

\begin{table*}[ht]
\centering
\caption{Reconstructed $M_B$ [mag] for all models, fiducials, and constraint configurations, PANTHEON dataset. Each model group shows two rows: DESI+CMB (top) and DESI+PP (bottom) fiducial. Bimodal cases list both peaks (peak$_1$\,/\,peak$_2$). Uncertainties are shown in Fig.~\ref{fig:mb_summary}.}
\label{tab:mb_hetero}
\begin{tabular}{lcccc}
\toprule
Model / Fiducial & H-TT & H-TF & H-FT & H-FF \\
\midrule
\textbf{$\Lambda$CDM} \\
\hspace{8pt}CMB & $-19.34$ & $-19.31$ & $-19.36\,/\,-19.31$ & $-19.38$ \\
\hspace{8pt}PP & $-19.30$ & $-19.34$ & $-19.31$ & $-19.37$ \\
\addlinespace[6pt]
\textbf{CPL} \\
\hspace{8pt}CMB & $-19.32$ & $-19.31$ & $-19.27$ & $-19.40\,/\,-19.29$ \\
\hspace{8pt}PP & $-19.34\,/\,-19.30$ & $-19.34$ & $-19.33$ & $-19.43$ \\
\addlinespace[6pt]
\textbf{GEDE} \\
\hspace{8pt}CMB & $-19.30\,/\,-19.27$ & $-19.34$ & $-19.37$ & $-19.35$ \\
\hspace{8pt}PP & $-19.30$ & $-19.33$ & $-19.33\,/\,-19.30$ & $-19.41$ \\
\addlinespace[6pt]
\textbf{$L_s$CDM} \\
\hspace{8pt}CMB & $-19.31$ & $-19.31$ & $-19.35$ & $-19.41$ \\
\hspace{8pt}PP & $-19.33$ & $-19.29$ & $-19.33$ & $-19.41$ \\
\bottomrule
\end{tabular}
\end{table*}

\begin{figure*}
    \centering
    \includegraphics[width=\textwidth]{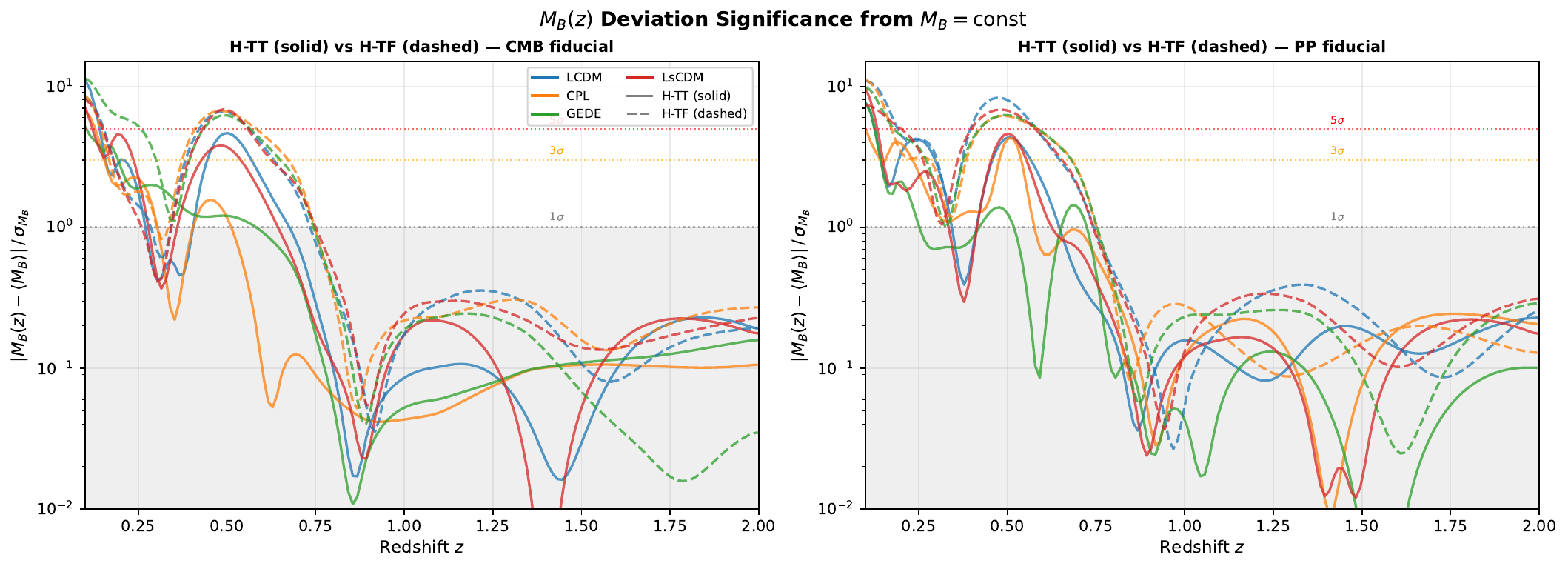}
    \caption{Significance of $M_B(z)$ deviation from its mean value
    $|M_B(z) - \langle M_B\rangle|/\sigma_{M_B}$, for H-TT (solid) and H-TF
    (dashed) configurations. Left panel: DESI+CMB fiducial; right panel:
    DESI+PP fiducial. Colors indicate cosmological model. Gray shading marks the
    $< 1\sigma$ region. Horizontal dotted lines indicate $1\sigma$, $3\sigma$,
    and $5\sigma$ thresholds. The uncertainty $\sigma_{M_B}$ is the
    observationally-floored standard error of the mean per redshift bin
    (Eq.~\ref{eq:sigma_floor}). The plot is restricted to $0.15 < z < 2$
    where SN~Ia coverage is reliable.}
    \label{fig:ddr_significance}
\end{figure*}
\paragraph{Statistical significance of $M_B$ evolution.}
\label{sec:significance}
To assess whether $M_B(z)$ deviates significantly from a constant fiducial value, we
compute the significance
\begin{equation}
S(z) = \frac{|M_B(z) - M_{B,\text{fid}}|}{\sigma_\mathrm{bin}(z)} ,
\end{equation}

where $\sigma_\mathrm{bin}(z)$ is the observationally-floored standard
error of the mean $M_B$ per redshift bin:
\begin{equation}
\sigma_{\rm bin}(z) = \frac{\max\!\left(\sigma_{\rm scatter}(z),\,
\bar{\sigma}_{\mu,\rm bin}\right)}{\sqrt{N_{\rm bin}(z)}},
\label{eq:sigma_floor}
\end{equation}
with $\sigma_{\rm scatter}$ the standard deviation of individual $M_B$
values within the bin and $\bar{\sigma}_{\mu,\rm bin}$ the mean
observational uncertainty of SNe~Ia in that bin.
This
quantity measures the statistical robustness of any $M_B(z)$ trend to the precision
actually achieved by the network, rather than the broader epistemic uncertainty on the
reconstruction itself.

Figure~\ref{fig:ddr_significance} shows the significance for the two
physically most constrained configurations (H-TT and H-TF) for the two fiducials in the reliable range $0.15 < z < 2$.

Under full constraints (H-TT) most models remain below $\sim 2\sigma$
in the well-sampled range $z \in [0.3, 1.5]$, consistent with a
constant $M_B$, though $\Lambda$CDM, CPL, and $\Lambda_s$CDM show
excursions above $3\sigma$ at $z \lesssim 0.3$ where BAO constraints
are absent and the low-$z$ SN~Ia systematics already discussed above. These deviations reoccur
at $z \sim 0.5$ for $\Lambda$CDM and $\Lambda_s$CDM under the CMB
fiducial, and for all models except GEDE under the PP fiducial.
H-TF (DDR only) shows $2$--$5\sigma$ peaks across $z \sim 0.4$--$0.75$
that are consistent across all four models and both fiducials,
suggesting a residual signal that the cosmological model prior
partially absorbs in H-TT. These features are robust to the choice
of fiducial: CMB and PP fiducial sets give qualitatively identical
significance profiles, with the PP fiducial showing somewhat lower
peak significance.

The partially and fully unconstrained configurations (H-FT and H-FF) reveal the role of each prior more starkly. Removing all constraints (H-FF) produces peaks of $3$--$5\sigma$ near $z \sim 0.4$--$0.5$, driven by network freedom rather than physical evolution --- H-FF is effectively an unconstrained ANN fit whose significance peaks should not be interpreted as a physical signal. H-FT (cosmology only, no DDR) is noisier and less consistent across models than H-TF, confirming that
DDR is the more essential constraint. The full comparison across all four configurations and both fiducials is shown in
Fig.~\ref{fig:ddr_significance_conservative} in Appendix~\ref{app:conservative_errors}.

The $z \sim 0.4$--$0.5$ peak corresponds to the densest SN~Ia coverage in Pantheon+, so it is driven by genuine data signal rather than sparse-sampling noise. Under a more conservative uncertainty treatment (Appendix~\ref{app:conservative_errors}) the peak significance is reduced by a factor of $\sim 2$, remaining above $1\sigma$ but falling below $3\sigma$. All models are consistent with $M_B = \mathrm{const}$ at
$z > 1.0$.

\paragraph{Low-Redshift Feature}
\label{sec:lowz}
All models and configurations show an elevated $M_B$ excursion at
$z < 0.1$, peaking near $z \sim 0.05$. This feature is consistent with
the low-$z$ spike reported in \cite{Benisty:2022psx} using Gaussian
Process methods and likely reflects the absence of BAO constraints below
$z \sim 0.1$ combined with the transition between local and Hubble-flow
SN~Ia subsamples. It is not interpreted as physical evolution since it happens in a region with no BAO coverage.

\begin{figure*}[!ht]
    \centering
    \includegraphics[width=\textwidth]{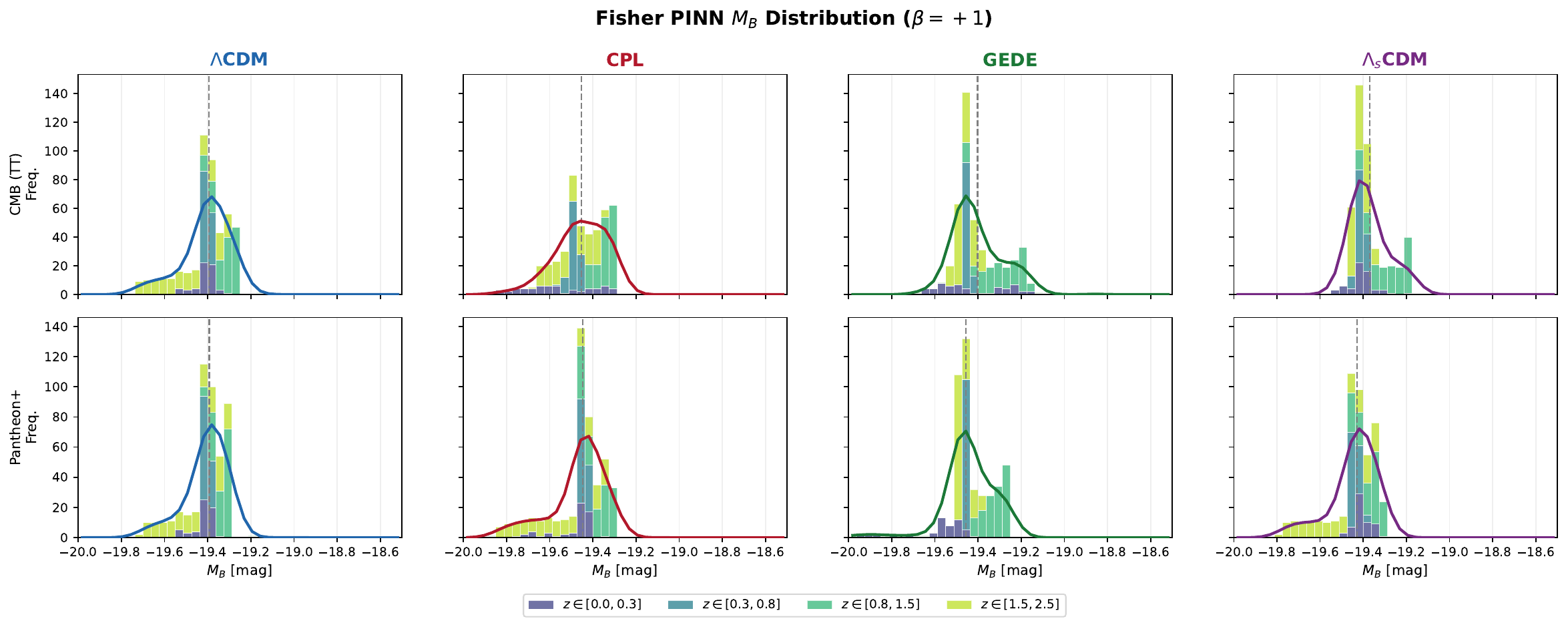}
   \caption{$M_B$ distributions reconstructed by the Fisher two-network method,
colour-coded by redshift bin ($z \in [0.0, 0.3]$, $[0.3, 0.8]$, $[0.8, 1.5]$,
$[1.5, 2.5]$). Top row: DESI+CMB fiducial; bottom row: Pantheon+ (DESI+PP)
fiducial.  GEDE and CPL show broader distributions reflecting additional dark energy freedom.
}
    \label{fig:fisher_distributions}
\end{figure*}

\subsection{Fisher Method}
\label{sec:fisher_results}

Figure~\ref{fig:fisher_distributions} presents the reconstructed $M_B$
distributions from the Fisher two-network method for all four cosmological
models and both fiducial priors, colour-coded by redshift bin. Several
features are noteworthy.

The redshift-binned distributions are clearly separated in all models:
low-$z$ bins ($z < 0.3$) consistently yield $M_B$ values $\sim 0.1$~mag
fainter than high-$z$ bins ($z > 0.8$), a pattern robust across all four
cosmologies and both fiducials. This separation confirms that the
two-network architecture successfully decouples the SN and BAO channels
--- any signal in $M_B(z)$ reflects a genuine tension between the two
datasets rather than internal self-consistency of a single fit.

The DESI+PP fiducial produces narrower distributions than DESI+CMB in all
models, consistent with the tighter constraint on the distance--redshift
relation provided by the local calibration. GEDE and CPL show broader
distributions than $\Lambda$CDM and $\Lambda_s$CDM, reflecting the
additional freedom in $w(z)$: models with evolving dark energy allow more
variation in $D_A(z)$, which propagates directly into the $M_B$ residual.

\begin{figure*}[!ht]
    \centering
    \includegraphics[width=0.45\textwidth]{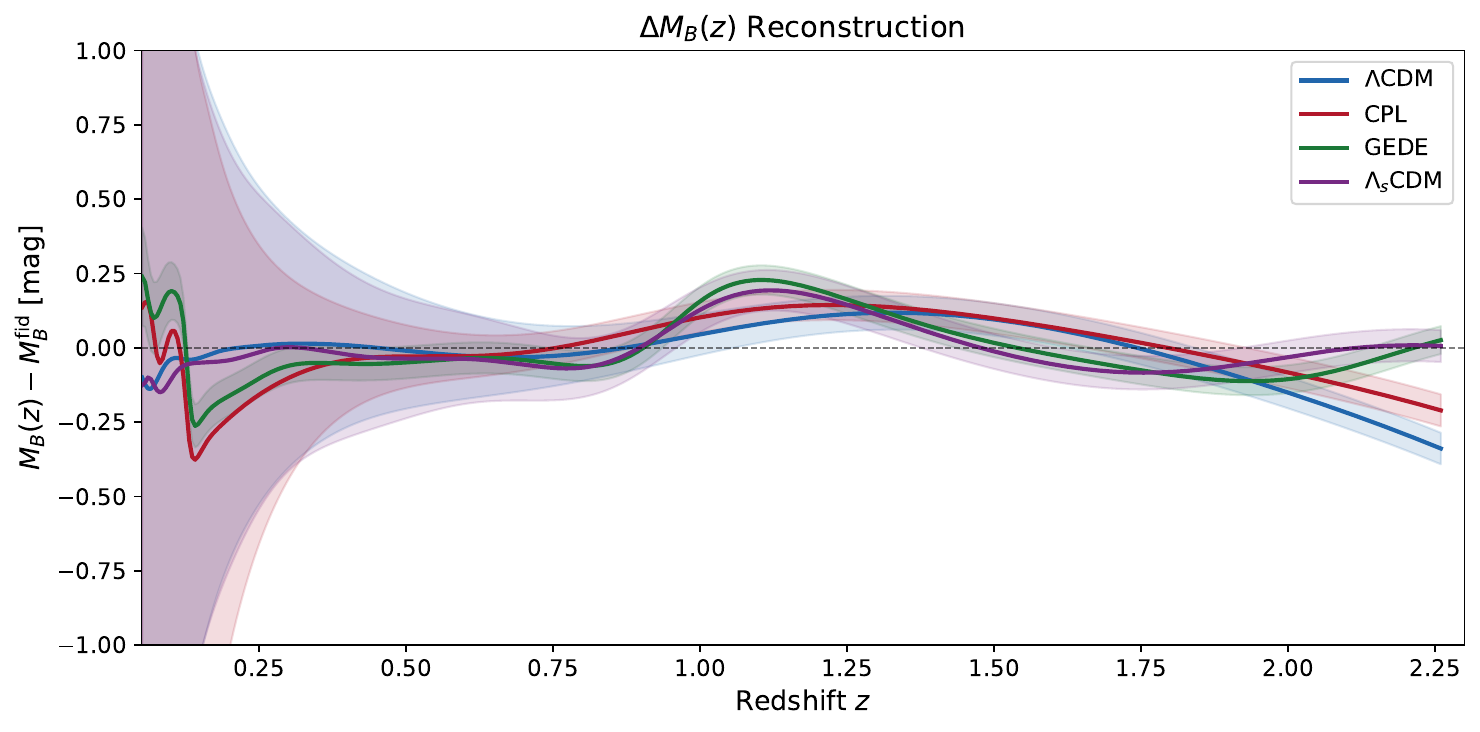}
    \includegraphics[width=0.45\textwidth]{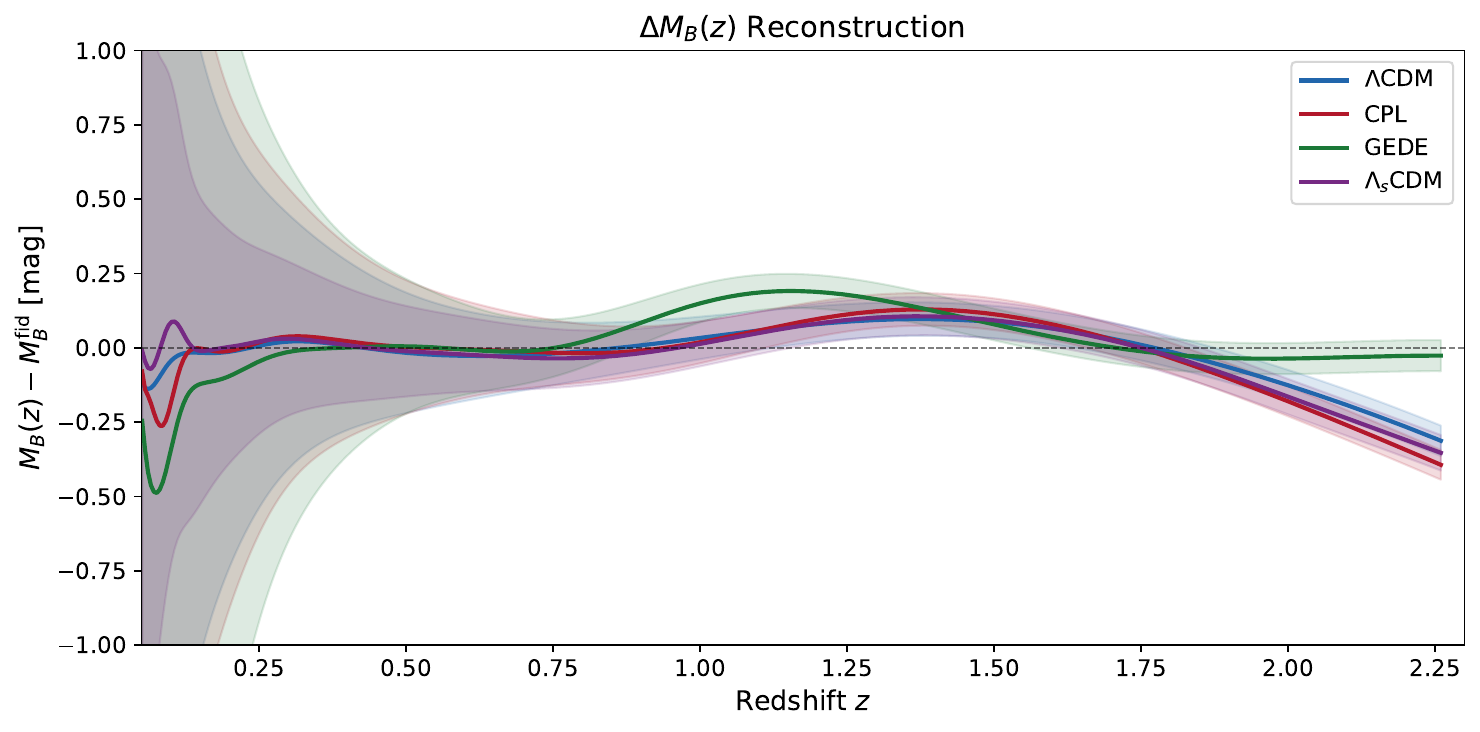}
    \caption{$\Delta M_B(z) = M_B(z) - M_B^\mathrm{fid}$ reconstructed by the
Fisher method. Left: DESI+CMB fiducial; right: DESI+PP fiducial. Shaded
bands show $1\sigma$ uncertainties. All models are consistent with zero
deviation in $z \in [0.3, 2]$. The oscillatory feature at $z \lesssim 0.2$
is data-driven and common to all models. At $z > 1.5$ models diverge as
SN coverage becomes sparse and reconstruction relies on BAO.}
    \label{fig:fisher_delta_MB}
\end{figure*}

Figure~\ref{fig:fisher_delta_MB} shows the continuous $\Delta M_B(z)$
reconstruction. In the well-constrained region $z \in [0.3,2]$ all
models are consistent with zero deviation at the $1\sigma$ level,
indicating no significant evidence for $M_B$ evolution in this range.
Notably, GEDE shows a positive excursion of $\sim 0.15$~mag around
$z \sim 1$ under the CMB fiducial that is absent under the PP fiducial,
suggesting this feature is driven by the tension between CMB-preferred
GEDE parameters and the BAO data rather than a genuine $M_B$ signal.
The oscillatory feature at $z \lesssim 0.2$ is common to all models and
both fiducials, suggesting a data-driven rather than model-dependent
origin, consistent with the low-$z$ feature already discussed.

At $z > 1.5$ the models diverge: $\Lambda$CDM shows a negative drift
of $\sim 0.3$~mag under the CMB fiducial that is absent in GEDE, while
the PP fiducial produces a uniform negative trend across all models at
$z > 2$. This high-$z$ behaviour should be interpreted with caution as
SN coverage is sparse and the reconstruction is driven primarily by
DA-net extrapolation beyond the last BAO measurement at $z \approx 2.33$.

The Fisher method's learned uncertainties grow appropriately at low and
high redshift, reaching $\sim 0.5$~mag at $z < 0.1$ where the
$\partial\mu/\partial D_A \propto 1/D_A$ divergence amplifies the DA-net
uncertainty into the $M_B$ error budget. This redshift-dependent
uncertainty structure represents an improvement over the heteroscedastic
method's tendency to underestimate uncertainty in sparse regions, as
discussed in Section~\ref{sec:uncertainty}.

Table~\ref{tab:fisher_MB} summarises the Fisher two-network reconstruction statistics.
The biases are systematically negative across all models and both fiducials, ranging from
$-0.07$~mag ($\Lambda_s$CDM, CMB) to $-0.16$~mag (GEDE, PP), indicating that the
independently reconstructed SN distances are consistently shorter than the BAO-inferred
distances at the fiducial $M_B = -19.3$~mag. These biases are larger than the
heteroscedastic values in Table~\ref{tab:mb_hetero} because the two-network architecture
eliminates the cross-talk through which the shared-backbone method partially absorbs the
SN--BAO tension into its fit; the Fisher biases therefore provide a more direct measure of
the offset between the two distance scales. The mean uncertainties
$\bar{\sigma}_{M_B} \sim 0.05$--$0.13$~mag in the well-constrained region $z \in [0.3, 2]$
are consistent with the observational error floor propagated through Eq.~(\ref{eq:sigma_MB}),
confirming that the inverse-Fisher NLL weighting produces physically motivated uncertainties
without artificial inflation.

\begin{table*}
\centering
\caption{Reconstructed $M_B$ statistics from the Fisher two-network
method, Pantheon+ dataset. Each model group shows two rows: DESI+CMB
(top) and DESI+PP (bottom) fiducial. The $N_\mathrm{peaks}$ column
shows 1 for unimodal distributions or the peak positions [mag] for
bimodal cases.}
\label{tab:fisher_MB}
\begin{tabular}{llcccccc}
\hline
Model & Fid. & Mean $M_B$ & $\sigma_\mathrm{dist}$ & Bias & $\bar{\sigma}_{M_B}$ & Bimodal & $N_\mathrm{peaks}$ \\
\hline
  $\Lambda$CDM & CMB & $-19.394$ & $0.105$ & $-0.094$ & $0.119$ & Yes & $-19.40/-19.31$ \\
   & PP & $-19.393$ & $0.092$ & $-0.093$ & $0.123$ & No & 1 \\
\hline
  CPL & CMB & $-19.453$ & $0.111$ & $-0.153$ & $0.080$ & Yes & $-19.48/-19.34$ \\
   & PP & $-19.447$ & $0.124$ & $-0.147$ & $0.132$ & No & 1 \\
\hline
  GEDE & CMB & $-19.402$ & $0.111$ & $-0.102$ & $0.051$ & No & 1 \\
   & PP & $-19.456$ & $0.116$ & $-0.156$ & $0.135$ & Yes & $-19.47/-19.30$ \\
\hline
  $\Lambda_s$CDM & CMB & $-19.370$ & $0.080$ & $-0.070$ & $0.125$ & No & 1 \\
   & PP & $-19.428$ & $0.106$ & $-0.128$ & $0.104$ & No & 1 \\
\hline
\end{tabular}
\end{table*}

\section{Discussion}
\label{sec:discussion}

\subsection{Hierarchy of Constraints}
\label{sec:discussion_hierarchy}

The systematic comparison across four constraint configurations establishes
that the Etherington distance duality relation provides more fundamental
regularisation than specific cosmological model assumptions. This is not
merely a numerical observation --- it reflects a deeper physical point:
the DDR is a geometric identity that holds in any metric theory of gravity
with photon number conservation, while the cosmological model prior encodes
assumptions about dark energy that may not be correct.

The extreme deviations of FT configurations (cosmology only, no DDR) are
instructive: without the geometric consistency enforced by the DDR, the
network is free to find solutions where the $D_A$ and $\mu$ heads are
mutually inconsistent, producing large DDR violations even when both heads
individually fit their respective data well. This suggests that for any
neural network approach to joint BAO+SN inference, DDR compliance should
be primary, regardless of what cosmological model
is assumed.

\subsection{Physical Interpretation and Model Comparison}
\label{sec:discussion_physics}

\textbf{Significance of $M_B(z)$ trends.}
The $2$--$3\sigma$ significance peaks at $z \sim 0.4$--$0.75$ seen in H-TF configurations and some TT ones reflect a genuine mild tension in the data, but require careful interpretation. With $\sim 100$--200 SNe~Ia per bin the uncertainty on the mean is $\sigma/\sqrt{N} \sim 10$--30~mmag, making even a 50~mmag systematic shift appear highly significant. That H-TT suppresses these peaks while H-TF does not indicates the cosmological model prior absorbs part of this tension — the prior
effectively pulls the reconstruction toward the assumed expansion history, masking residuals that appear when only the DDR is enforced. The feature at $z \sim 0.4$--$0.5$ is present in both fiducials and all four models, ruling out a model-specific or fiducial-specific artifact. That it persists even under H-TT for some models suggests the tested cosmologies cannot fully absorb this tension. For reference, a shift of $\sim 50$ mmag in $M_B$ corresponds to a $\sim 2\%$ change in the inferred luminosity distance, placing the effect at a level directly relevant for current precision cosmology constraints.

\textbf{Fisher method and fiducial dependence.}
The Fisher method provides a complementary and more direct probe of
this tension. Under full constraints it finds no significant deviation
from $M_B = \mathrm{const}$ in $z \in [0.3, 0.75]$, but shows a
minor excursion around $z \sim 1$ that is more pronounced for GEDE under the CMB fiducial and nearly absent under the PP fiducial. This fiducial dependence is physically meaningful: DESI+CMB and DESI+PP correspond to different underlying datasets and preferred parameter combinations, so a feature that appears under one but not the other
points to an incompatibility between the CMB-preferred parameters and the joint BAO+SN dataset at these redshifts, likely related to the known tension between the PP and CMB datasets, rather than a genuine
$M_B$ signal. The PP fiducial produces a nearly flat $\Delta M_B(z)$ for all models in the well-constrained region, while the CMB fiducial shows a broader excursion peaking near $z \sim 1$ — the same redshift where the heteroscedastic method shows its secondary feature, though
shifted slightly from the $z \sim 0.4$--$0.5$ primary peak. Whether this shift reflects a genuine difference between the two method architectures or a dependence on the uncertainty floor deserves further investigation.

\textbf{Model comparison and bimodality.}
$\Lambda$CDM and $\Lambda_s$CDM show the most stable $M_B$
reconstruction across all configurations, remaining unimodal and
within 0.05~mag of $-19.3$ even in partially constrained runs.
GEDE and CPL show the strongest evidence for tension, exhibiting
bimodality under TT and larger scatter in the significance profiles,
consistent with their additional dark energy freedom allowing more
degenerate solutions in the $D_A$--$\mu$ loss landscape. For the
heteroscedastic method this bimodality reflects the network finding
two equally valid fits to the data; for the Fisher method it reflects
an incompatibility between the data and the assumed fiducial
parameters. In both cases the practical implication is the same:
analyses inferring $H_0$ from SN data alone without BAO anchoring
may encounter multimodal posteriors for CPL and GEDE, and single-mode
summaries of such posteriors would be misleading.

\textbf{Comparison with independent $M_B$ calibrations.}
The recovered $M_B \approx -19.3$~mag is consistent with 
late-universe calibrations: the Tip of the Red Giant Branch 
(TRGB) yields $M_B \sim -19.26$~mag ~\cite{Freedman:2024eph, Li:2025lfp} and the Cepheid-based 
SH0ES calibration gives $M_B \sim -19.25$~mag~\cite{Riess:2024vfa}. The Planck-preferred value of $-19.44$~mag lies 
$\sim 0.14$~mag away, consistent with the constant offset 
introduced by our choice of $r_d = 147.0$~Mpc (see 
Section~\ref{subsec:Data}): as shown in~\cite{Benisty:2022psx}, 
alternative $r_d$ values shift $M_B$ by a constant without 
affecting the redshift-dependent reconstruction. The 
agreement with late-universe calibrations therefore reflects 
agreement with our $r_d$ prior rather than an independent constraint on 
the distance ladder.

\subsection{Comparison with Previous Work}
\label{sec:discussion_gp}

Our results are broadly consistent with the GP and ANN reconstructions of
\cite{Benisty:2022psx} and \cite{Staicova:2024dak}, which found $M_B$
constant within $1\sigma$ at current data precision. The PINN approach
adds two capabilities those methods lack: explicit enforcement of physical
constraints through loss terms, yielding DDR compliance of 30--50~mmag
even in sparse data regions where GP relies entirely on kernel smoothness;
and the two-network Fisher architecture which ensures that any $M_B(z)$
signal reflects a genuine tension between independent BAO and SN datasets
rather than a fitting artefact.

The heteroscedastic uncertainties ($\sigma \sim 0.13$--0.15~mag for the
primary variant) are smaller than typical GP estimates ($\sim 0.15$--0.25~mag)
because the NLL formulation incorporates epistemic model uncertainty, while
GP uncertainties reflect only kernel-based interpolation variance. As
discussed in Section~\ref{sec:uncertainty}, neither approach is strictly
superior, they measure different things --- the PINN uncertainty reflects what the data can accommodate, while the GP uncertainty reflects what the model cannot exclude. The conservative excess-variance variant
(Appendix~\ref{app:conservative_errors}) produces uncertainties of
$\sim 0.2$--0.7~mag, closer to GP estimates, while leaving the mean
$M_B(z)$ reconstruction unchanged to within 5~mmag.

Taken together, the two PINN architectures provide a consistent picture of the observed feature. In the heteroscedastic framework it appears as a localized $2$--$3\sigma$ deviation in $M_B(z)$, while in the Fisher two-network construction it manifests as a systematic separation between redshift-binned $M_B$ distributions. Despite these differing representations, both approaches point to a mild but persistent tension between the BAO- and SN-inferred distance scales localized at $z \sim 0.4$--$0.5$. This consistency across architecturally distinct methods supports the interpretation of the feature as data-driven rather than a byproduct of a specific network design and warrants consideration as either unaccounted
systematic effects in the SN~Ia standardisation at these redshifts or
a mild signal of new physics in the distance--redshift relation.
\subsection{Implications for Cosmological Inference}

A redshift-dependent shift in the inferred absolute magnitude of Type Ia supernovae directly propagates into cosmological parameter estimation through the distance ladder. A luminosity distance shift of $\mathcal{O}(2\%)$ is comparable to the precision of current late-universe probes. If interpreted as a genuine evolution in $M_B(z)$, such a shift would bias the inferred value of $H_0$ and any other parameter constrained through supernova distance measurements. Alternatively, if the observed feature reflects residual systematics in the supernova standardisation, it would indicate that current calibration procedures may not fully capture redshift-dependent effects in the range $z \sim 0.4$--$0.5$. In either case, the observed localized deviation is sufficient to affect precision cosmology analyses and motivates further investigation with upcoming datasets.

\section{Conclusions}
\label{sec:conclusions}

We have applied two variants of Physics-Informed Neural Networks to
reconstruct $M_B(z)$ from joint BAO and SN~Ia data under four
cosmological models and two DESI~DR2 fiducial sets.  A key result of this work is the identification of a hierarchy of constraints: enforcing the Etherington distance duality relation yields substantially more stable and physically consistent reconstructions than imposing cosmological model priors alone. This suggests that geometric consistency conditions should be treated as primary constraints in machine-learning-based cosmological inference frameworks. Concretely, the heteroscedastic method establishes that enforcing the Etherington DDR
yields $2$--$50\times$ smaller violations than enforcing only the cosmological model prior.  Under full constraints all models recover $M_B \approx -19.3$~mag,
consistent with standard supernova standardisation at current
precision.

The Fisher two-network method, training independent networks on each
probe, finds no significant $M_B$ evolution in $z \in [0.3, 1.0]$
under full constraints. The bimodality of $M_B$ distributions for
CPL and GEDE in unconstrained runs is a novel result: analyses
inferring $H_0$ from SN data alone without DDR enforcement may
encounter multimodal posteriors for these models.

While the feature appears consistently across models and fiducial choices, its origin remains uncertain. Its persistence across distinct reconstruction strategies suggests it reflects a property of the joint BAO+SN dataset rather than a methodological artifact, but the reasons behind it may differ. In particular, redshift-dependent observational systematics or population effects in the supernova sample could produce a similar signal that is largely insensitive to the assumed cosmological model. We therefore interpret the observed behaviour as consistent with either residual systematics or a mild physical deviation in the distance--redshift relation, and defer definitive conclusions to future data, such as the forthcoming DESI data releases and next-generation SN~Ia
surveys.

\section*{Acknowledgement}
This research was funded by Bulgarian National Science Fund grant number KP-06-N88/1.

\appendix

\section{Fiducial Parameters and Implementation Details}
\label{app:implementation}

Table~\ref{tab:models} lists the fiducial cosmological parameters used
in this work. Both PINN variants use the Adam optimiser \cite{Kingma:2014vow}
with learning rate $10^{-3}$ and gradient clipping (max norm 1.0). The
random seed is fixed (\texttt{SEED}=43) for reproducibility. Implementation
uses TensorFlow~2.x and SciPy.

\begin{table*}[h]
\centering
\caption{Fiducial cosmological parameters as published in
\cite{DESI:2025zgx} for $\Lambda$CDM and CPL, \cite{Chaudhary:2025vzy}
for GEDE, and \cite{Akarsu:2024eoo} for LsCDM. All models assume
$\Omega_k = 0$.}
\label{tab:models}
\begin{tabular}{lll}
\hline
Model & DESI+CMB & DESI+PP \\
\hline
$\Lambda$CDM
  & $\Omega_m = 0.3027$
  & $\Omega_m = 0.2975$ \\[0.25cm]
CPL
  & $\Omega_m=0.352$, $w_0=-0.43$, $w_a=-1.70$
  & $\Omega_m=0.298$, $w_0=-0.888$, $w_a=-0.477$ \\[0.25cm]
GEDE
  & $\Omega_m=0.306$, $\Delta=-1.10$
  & $\Omega_m=0.306$, $\Delta=-0.94$ \\[0.25cm]
LsCDM
  & $\Omega_m=0.300$, $z_\dagger=2.0$
  & $\Omega_m=0.295$, $z_\dagger=1.8$ \\
\hline
\hline
\end{tabular}
\end{table*}

\paragraph{Heteroscedastic method.} Training proceeds in three phases
of progressive constraint ramping: Phase~1 (3000 steps,
$0.1\times\lambda_\mathrm{target}$) establishes the data fit; Phase~2
(5000 steps, $0.5\times\lambda_\mathrm{target}$) introduces physics;
Phase~3 (8000 steps, $\lambda_\mathrm{target}$) enforces full
constraints. Collocation points are resampled uniformly at each step.

\paragraph{Fisher method.} Training proceeds in four phases. A warmup
(1000 steps, MSE only) initialises both networks before any NLL or
physics terms are active. Phase~1 (2000 steps,
$0.1\times\lambda_\mathrm{phys}$) and Phase~2 (9000 steps, full
$\lambda_\mathrm{phys}$) both use stop-gradient on the residuals fed
to the variance NLL. Phase~3 (3000 steps) removes the stop-gradient,
reduces the learning rate to $0.3\times$, tightens gradient clipping
to max norm 5, and reduces physics regularisation
($\lambda_\mathrm{phys} \to 0.5\lambda_\mathrm{phys}$,
$\lambda_\mathrm{deriv} \to 0.3\lambda_\mathrm{deriv}$).

\paragraph{Regularisation terms.} Both networks penalise curvature
of their mean output and log-variance output on a uniform grid of
60 points spanning $z \in [0.001, 2.5]$:
\begin{equation}
\mathcal{L}_\mathrm{reg} = \lambda_\mathrm{sm}
  \sum_k (f_{k+1} - 2f_k + f_{k-1})^2
  + \lambda_\mathrm{sv}
  \sum_k (s_{k+1} - 2s_k + s_{k-1})^2,
\end{equation}
where $f$ is the mean output ($D_A/r_d$ or $\mu$), $s$ is the
log-variance output, and $\lambda_\mathrm{sm} = 10$,
$\lambda_\mathrm{sv} = 50$. The DA-net regularisation additionally includes a boundary loss
$\mathcal{L}_\mathrm{bnd}$ pinning DA-net to the theoretical $D_A/r_d$
value at $z = 0.001$ and $z = 2.5$, with non-finite theory values
masked out for non-standard cosmologies.

\section{Cross-Validation and Outlier Analysis}
\label{app:crossval}

To test generalisation we perform cosmic $K$-fold cross-validation by
dividing the redshift range into $K$ windows, training on all data
outside each window, and measuring prediction smoothness inside via
the standard deviation of the second finite difference:
\begin{equation}
S = \mathrm{std}\!\left[f(z_{i+2}) - 2f(z_{i+1}) + f(z_i)\right].
\end{equation}
Lower $S$ indicates smoother extrapolation. All methods degrade at
$z < 0.3$ and $z > 1.5$; H-FF shows the highest smoothness variation,
consistent with overfitting risk in the absence of constraints.

Outliers are identified via standardised residuals
$r_i = (y_i - y_\mathrm{theory})/\sigma_i$, flagging $|r_i| > 3$.
Two BAO outliers are found at $z = 0.51$ and $z = 0.93$ across all
models, likely reflecting systematic effects in those survey
measurements. No SN~Ia outliers are found at the $3\sigma$ level;
the covariance matrix adequately accounts for correlated systematics.

\section{Conservative Heteroscedastic Loss}
\label{app:conservative_errors}

The standard heteroscedastic loss with $w = 0.1$ weighting on the
NLL uncertainty term causes the network to learn
$\sigma_\mathrm{model} \approx 0$, yielding per-bin scatter
systematically below the observational noise floor (ratio 0.34--0.92
across bins). To address this we implement an excess-variance
parametrisation in which the network output $s(z)$ encodes the
excess variance above the observational floor:
\begin{equation}
\sigma_\mathrm{total}^2(z) = \sigma_\mathrm{obs}^2(z)
  \left(1 + e^{s(z)}\right),
\label{eq:excess_var}
\end{equation}
guaranteeing $\sigma_\mathrm{total} \geq \sigma_\mathrm{obs}$
at all redshifts. The corresponding loss
\begin{equation}
\mathcal{L}_\mathrm{SN}^\mathrm{cons} = \frac{\chi^2_\mathrm{SN}}{N_\mathrm{SN}}
  + \left\langle \frac{r_i^2}{\sigma_{\mathrm{obs},i}^2(1+e^{s_i})}
  + \frac{1}{2}\log(1+e^{s_i}) \right\rangle
\label{eq:loss_excess}
\end{equation}
is bounded below by zero, with minimum at
$e^s = r_i^2/\sigma_{\mathrm{obs},i}^2 - 1$ when residuals exceed
the observational floor and at $s \to -\infty$ otherwise.

\begin{figure*}
\centering
\includegraphics[width=2\columnwidth]{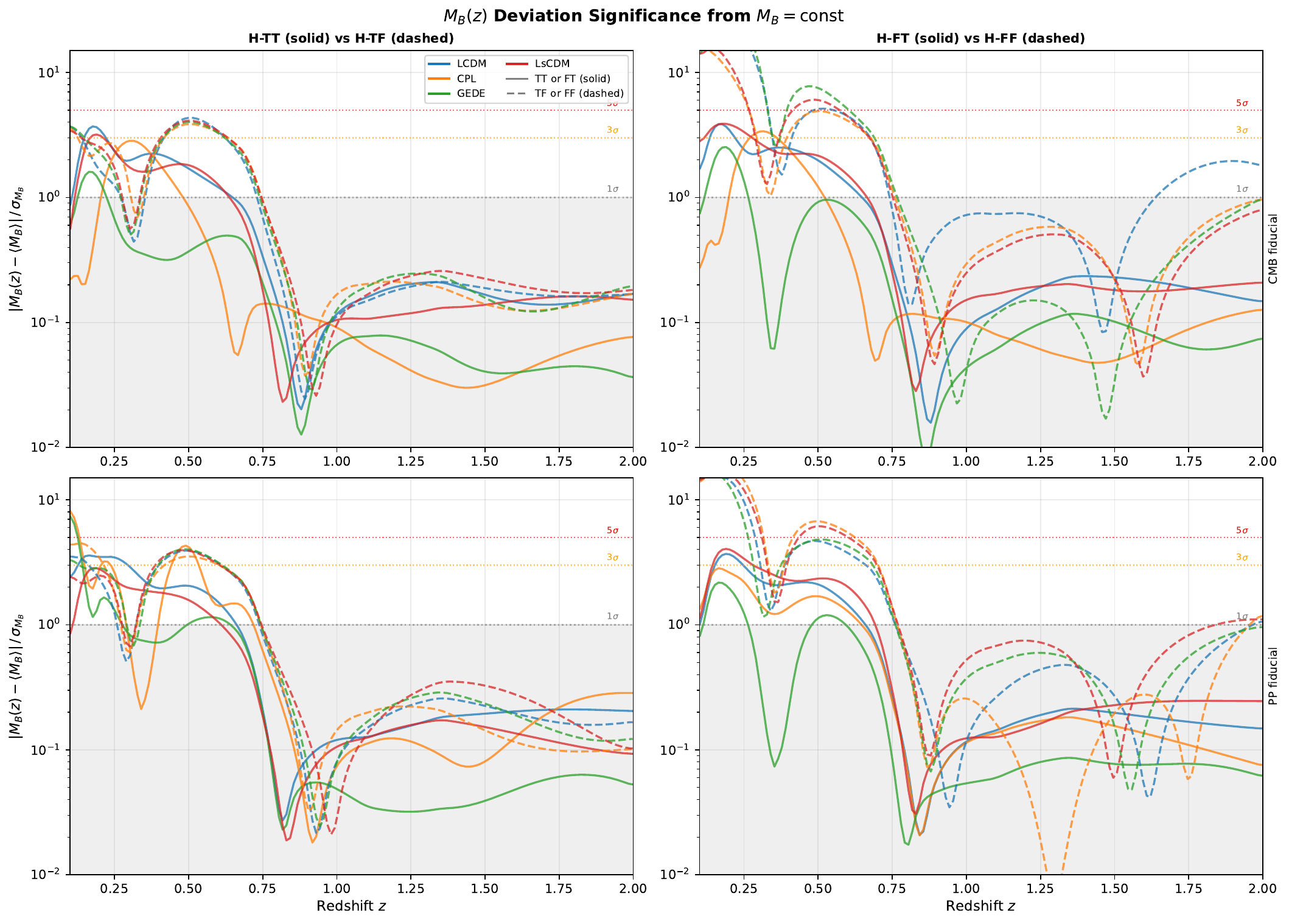}
\caption{$M_B(z)$ deviation significance for all four constraint
configurations (H-TT, H-TF, H-FT, H-FF) and both fiducials using the
excess-variance loss (Eq.~\ref{eq:loss_excess}). The $z \sim 0.4$--0.5
significance peak is reproduced in both constrained configurations,
confirming robustness to the uncertainty parametrisation.}
\label{fig:ddr_significance_conservative}
\end{figure*}

Rerunning the full $\Lambda$CDM analysis with this loss for all four
constraint configurations confirms that the mean $M_B(z)$ curves are
indistinguishable from the standard-loss runs (differences $< 5$~mmag),
while uncertainties increase by a factor of $\sim 10$--30 to
$\sigma_\mathrm{total} \sim 0.2$--0.7~mag. The $z \sim 0.4$--0.5
significance peak in Fig.~\ref{fig:ddr_significance_conservative}
is reproduced at comparable significance, confirming that the main
conclusions are robust to the choice of uncertainty parametrisation.

\end{document}